\documentclass[twocolumn,twoside,slac_two]{revtex4}
\usepackage{graphicx}
\usepackage{fancyhdr}
\newcommand{\noi}{\noindent}

\begin{document}

\title{Blazar observations above 100\,GeV with VERITAS}

%

\author{M. Errando}
\affiliation{Department of Physics \& Astronomy, Barnard College, Columbia University, 3009 Broadway, New York, NY 10027, USA}
\author{for the VERITAS Collaboration}

\begin{abstract}
The VERITAS array of 12-m atmospheric-Cherenkov telescopes in southern Arizona is one of the world's most-sensitive detectors of very-high-energy (VHE, $E>$100 GeV) gamma rays. More than 50 extragalactic sources are known to emit VHE photons; these include blazars, radio galaxies, and starburst galaxies. Blazar observations are one of the VERITAS Collaboration's Key Science Projects. More than 400 hours per year are devoted to this program and $\sim 100$ blazars have already been observed with the array, in most cases with the deepest ever VHE exposure. These observations have resulted in 21 detections, including 10 VHE discoveries, all of them with supporting multiwavelength observations. Recent highlights from VERITAS extragalactic observation program and the collaboration's long-term blazar observation strategy are presented.
\end{abstract}

\maketitle

\thispagestyle{fancy}

\section{The population of VHE blazars}
Blazars are active galactic nuclei (AGN) hosting a supermassive black hole, with relativistic jets pointing close to the line of sight to the observer. 
The small viewing angle of the jet makes it possible to observe strong relativistic effects, such as a boosting of the emitted power and shortening of the characteristic time scales. 
Blazars are typically identified in radio and optical surveys, their defining characteristics being radio loudness, flat spectrum and compact morphology in the radio band, some degree of optical polarization, and fast, large-amplitude variability.

\begin{figure*}[t]
\centering
\includegraphics[height=2.4in]{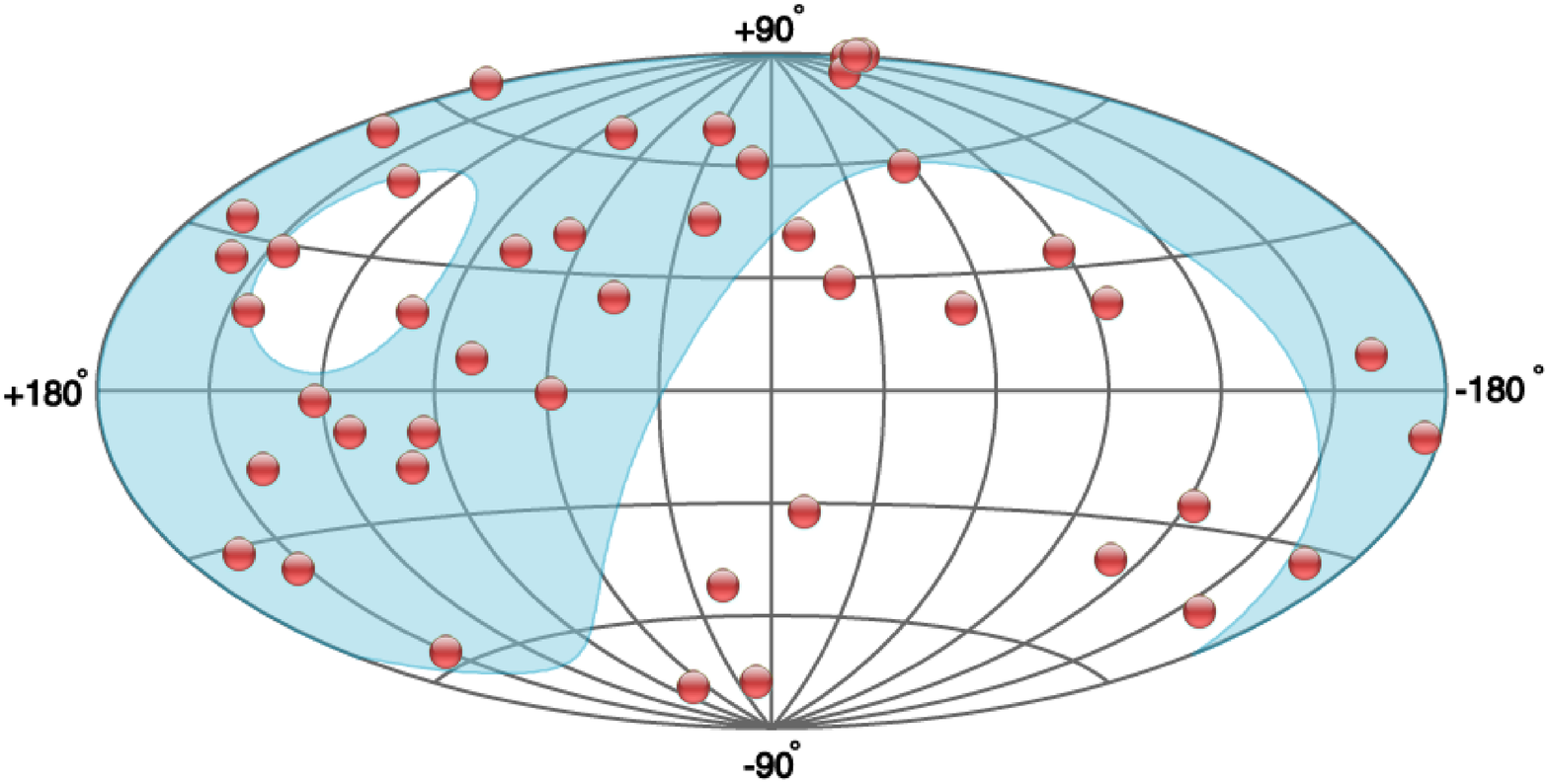}
\includegraphics[height=2.4in]{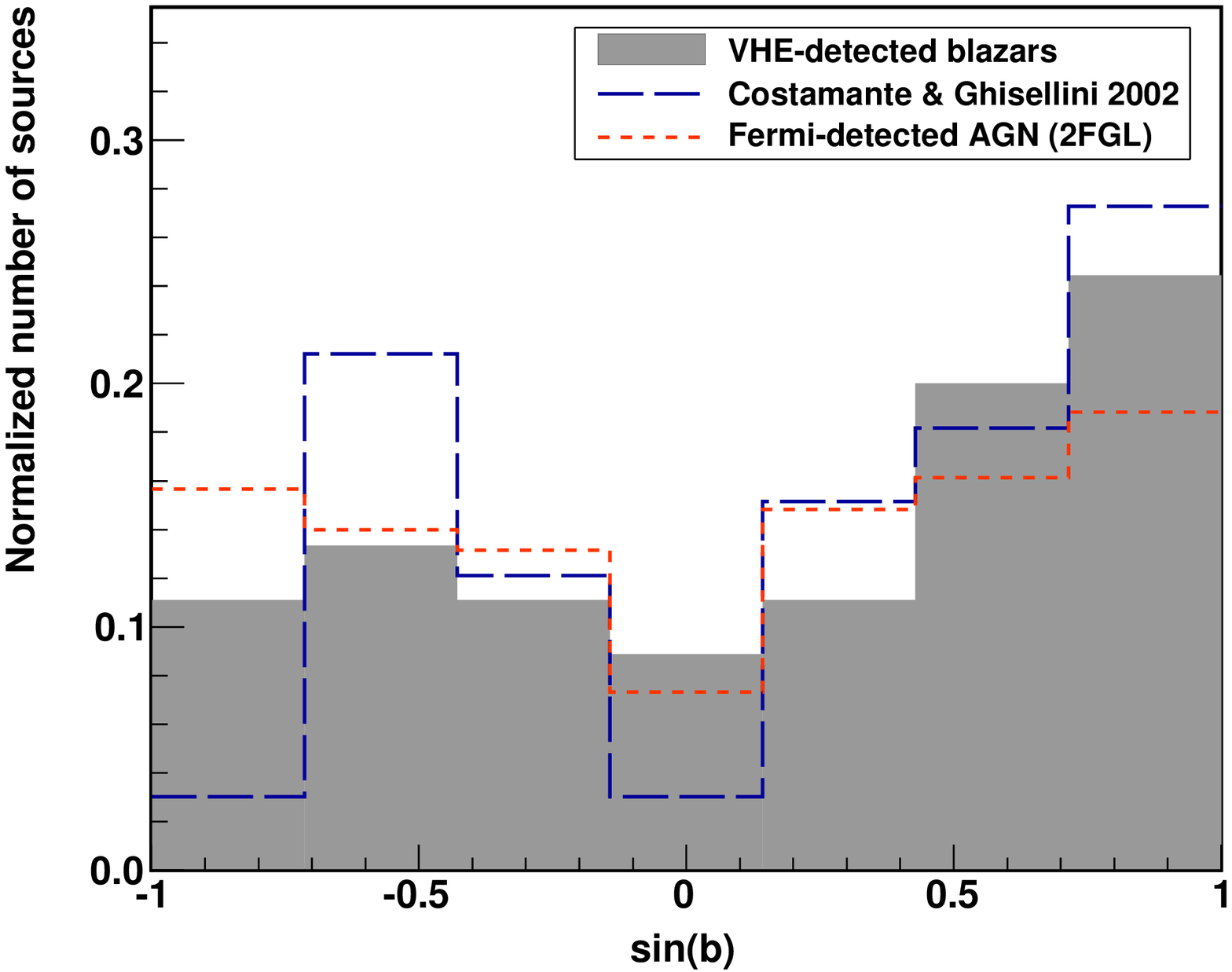}
\hspace{1cm}
\includegraphics[height=2.4in]{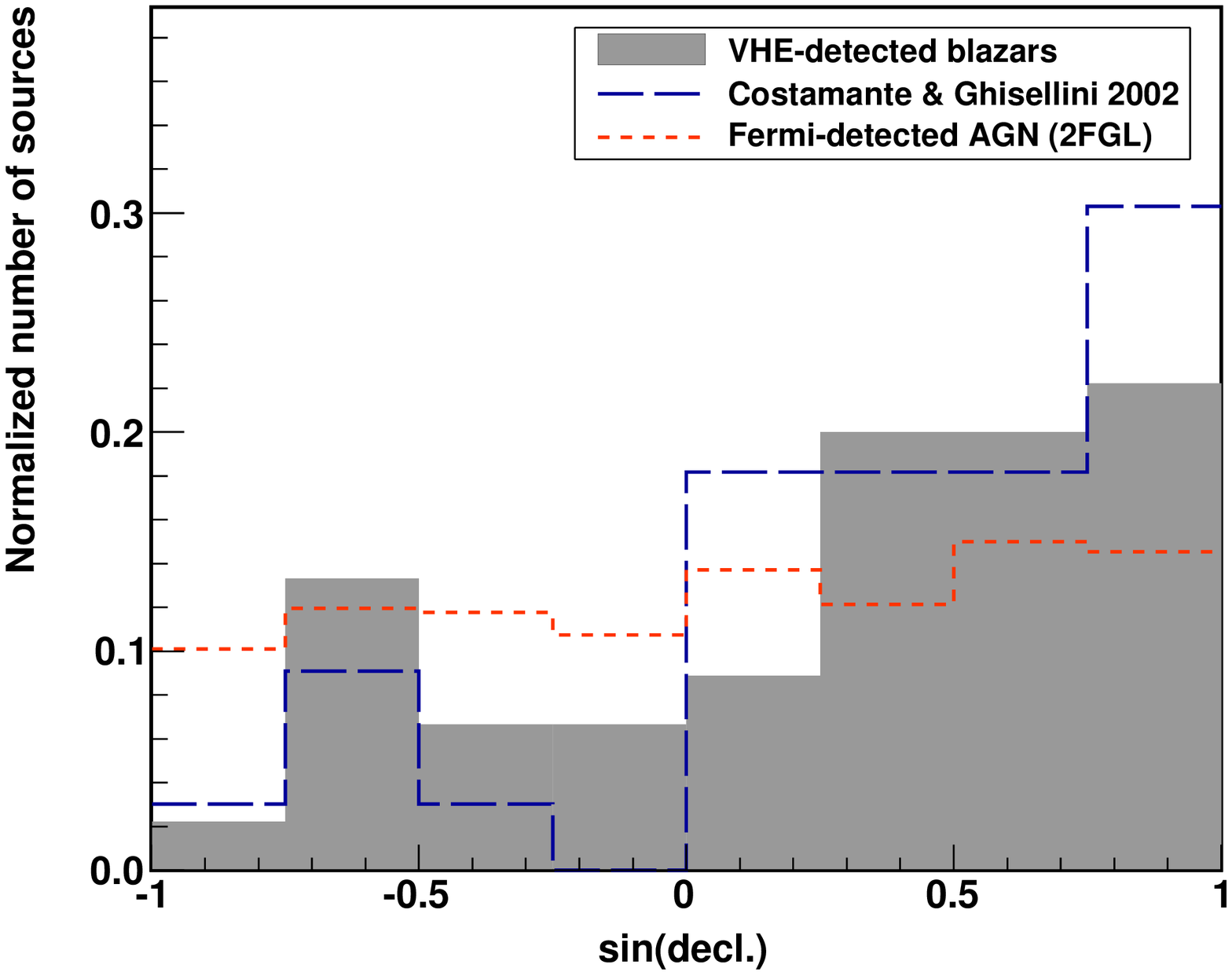}
\caption{{\em Top:} All-sky map of all VHE-detected blazars in galactic coordinates from \cite{tevcat}. The shaded area indicates the region observable with VERITAS. {\em Bottom:} Distribution of VHE blazars in galactic latitude ({\em left}) and celestial declination ({\em right}). The corrsponding distributions are shown for candidate VHE blazars \cite{costamante} and {\em Fermi}-detected blazars \cite{2fgl} for comparison.} 
\label{fig:map}
\end{figure*}

\begin{figure}[t]
\centering
\includegraphics[width=0.9\columnwidth]{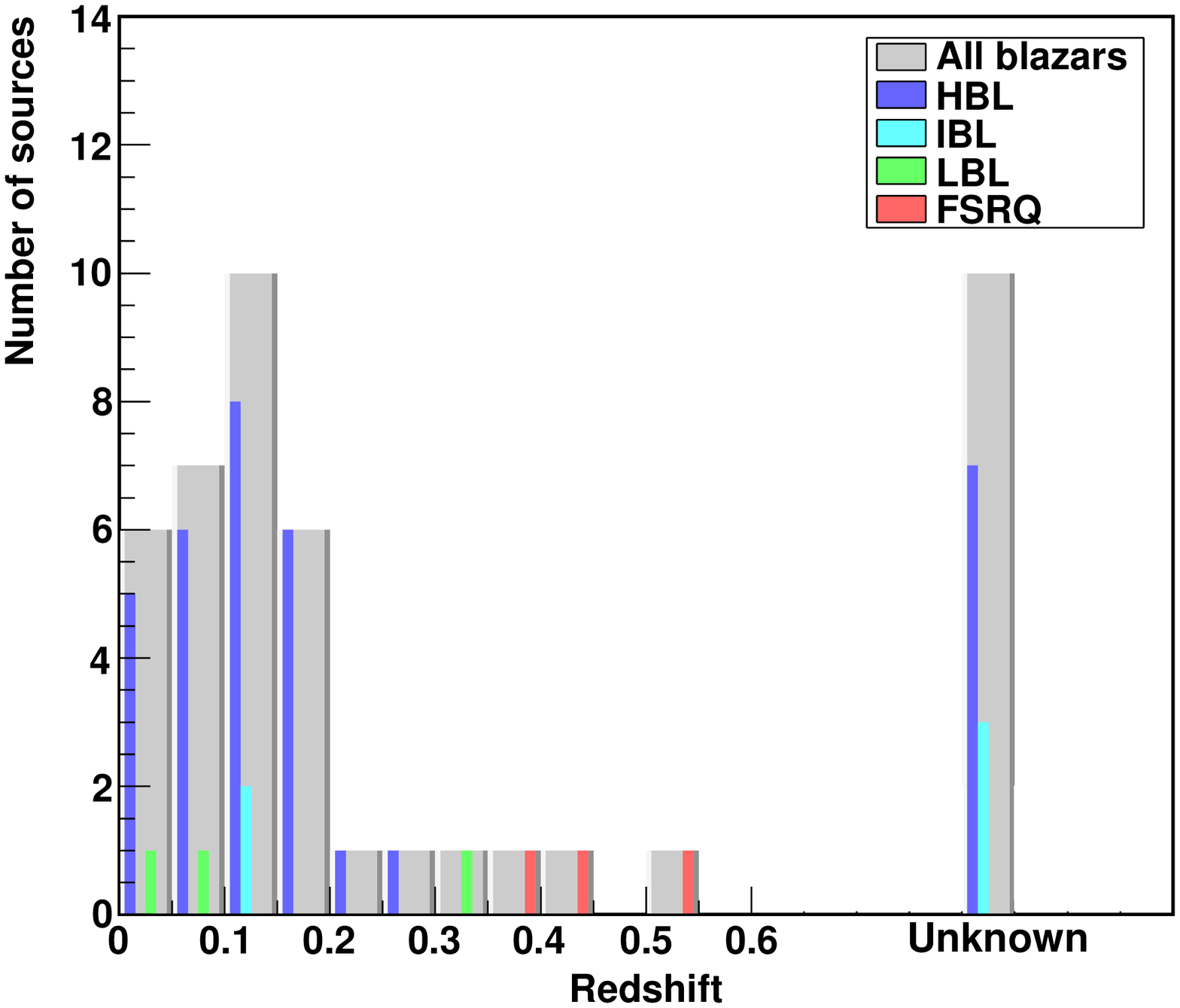}
\caption{Redshift distribution of all VHE-detected blazars, including the distribution for each blazar class. Data taken from \cite{tevcat}. Blazars with unknown redshift include objects without a quoted redshift in \cite{tevcat} 
and blazars with published redshift values that have subsequently been reported as uncertain.
} 
\label{fig:z}
\end{figure}

Since the first detection of a blazar in the very-high-energy gamma-ray band (VHE, $E>100$\,GeV) in 1992 by the Whipple Collaboration \cite{421-whipple}, ground-based imaging Cherenkov telescopes have discovered up to 45 VHE-emitting blazars \cite{tevcat}. 
This number is growing fast, with half of these detections reported in the last three years. VHE blazars populate the whole sky. 
Figure~\ref{fig:map} shows the distribution of VHE blazars in galactic latitude and declination. 
The only noticeable asymmetry is that two thirds of the sources are located in the northern half of the sky.
In the redshift distribution of VHE blazars (see Figure~\ref{fig:z}) most objects occur at $z<0.2$, with the most distant known VHE source being 3C~279 at $z=0.536$ \cite{magic-279}. 
The pair-production of VHE photons with UV to infrared photons from the extragalactic background light (EBL) limits the transparency window for gamma-rays at large redshifts. 
Finally, a significant fraction of VHE blazars (10 out of 45) have unknown redshift, or published redshifts that have been reported as uncertain. 

\begin{table}[b]
\begin{center}
\begin{tabular}{c  c c c c c c }
\hline
\hline
Class & \# of sources & $\langle z \rangle$$^a$ & & $\langle\Gamma\rangle$$^b$ & & $t_{var}$$^c$ \\
\hline
HBL & 33 & 0.12 & &3.2 & & $<5$\,min, \cite{magic-501,hess-2155}   \\
IBL & 5 & 0.12 & &3.7 & & $\sim 1$\,day, \cite{ver-66a-1}\\
LBL & 4 & 0.14 & &3.2 & & $<20$\,min, \cite{icrc-bllac}\\
FSRQ & 3 & 0.44 & & 3.9 & & $\sim 10$\,min, \cite{magic-1222} \\
Total & 45 & & \\
\hline\hline

\end{tabular}
\caption{Average properties of all VHE-detected blazars divided by class: high, intermediate, and low-frequency-peaced BL~Lacs (HBL, IBL, LBL, respectively) and flat spectrum radio quasars (FSRQ). Data taken from \cite{tevcat}. \\$^a$ Average redshift, excluding the sources where the redshift is unknown or uncertain. \\$^b$ Average photon index in the VHE range. \\$^c$ Shortest observed variability time scale in the VHE band, with the corresponding reference.}\label{tev_table}
\end{center}
\vspace{-0.3cm}
\end{table}

\begin{figure}[t]
\centering
\includegraphics[width=0.99\columnwidth]{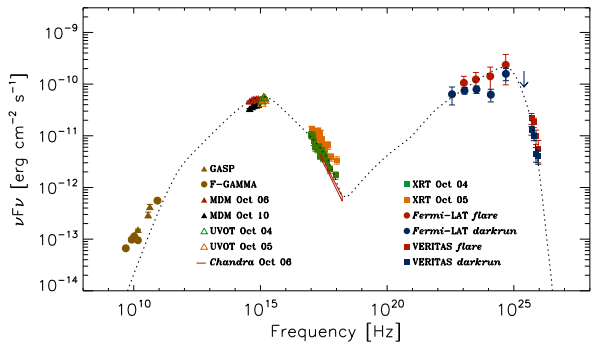}
\caption{Broadband SED of 3C 66A during the 2008 October multiwavelength campaign \cite{ver-66a-2}. The dotted line shows an SSC+EC model assuming a redshift of $z=0.3$.} 
\label{fig:66a}
\end{figure}

Blazars display a characteristic two-component spectral energy distribution (SED, see Figure~\ref{fig:66a}). 
The low-frequency continuum is usually attributed to synchrotron emission by relativistic electrons inside the jet magnetic field. 
VHE observations of blazars sample the high-frequency peak of the SED. 
Most theoretical models assume that the gamma-ray emission comes from inverse-Compton (IC) scattering of low-energy photons off of a population of relativistic electrons.
The source of photons for the IC process can be either synchrotron photons emitted by the same electron population \cite[so-called synchrotron self-Compton models (SSC), e.g.,][]{maraschi}, or a population of photons external to the jet \cite[external Compton (EC) models, e.g.,][]{dermer93}. 
Another family of models includes emission from hadrons in the jet \cite[e.g.,][]{mannheim}, establishing AGN jets as candidates for the origin of the highest energy cosmic rays \cite[see][]{kotera-olinto}.

Table~\ref{tev_table} shows the average properties of VHE-detected blazars divided by blazar class. Most detected sources are of the BL~Lac class, with the only exception of three flat-spectrum radio quasars (FSRQ). VHE-emitting FSRQs exhibit very soft spectral index and are only detected during very bright flaring episodes \cite{magic-279,magic-1222}. FSRQs are also the most distant sources detected in the VHE range. High-frequency-peaked BL~Lacs (HBL) are the dominant subclass of VHE-detected blazars, accounting for three quarters of the total source count. HBLs are defined by having the peak of the low-energy (or synchrotron) component ($\nu_{sync}$) at frequencies higher than $10^{15}$\,Hz.
Observations of hard-spectrum, mid-redshift HBLs have been used in recent years to place strong constraints on the density of the EBL \cite{hess-ebl,hess-0229}, to study the formation of pair cascades in the propagation of TeV photons through the intergalactic medium and to derive constraints on the strength of intergalactic magnetic fields \cite{igmf-neronov,igmf-tavecchio,igmf-dermer}.
Intermediate-frequency-peaked BL~Lacs (IBL, $10^{14}\,\mathrm{Hz} < \nu_{sync}< 10^{15}\,\mathrm{Hz}$) and low-frequency-peaked BL~Lacs (LBL, $\nu_{sync}< 10^{14}\,\mathrm{Hz}$) are typically detected during high flux states, with low gamma-ray states falling at the level or below the sensitivity of current instruments. The SED of VHE-detected IBLs and LBLs often  shows evidence for a significant contribution of EC processes to the total gamma-ray output \cite[e.g.,][see Figure~\ref{fig:66a}]{ver-wcom-2,ver-66a-2}.

The fastest flux variability in blazars has been observed in the VHE range, with individuals of most blazar subclasses exhibiting sub-hour flux variations (see Table~\ref{tev_table}). The observed flux-doubling times indicate compact emission regions with size $<10^{-3}$\,pc \cite{begelman}. Such compact regions suggest that the emission could originate from the base of the jet, in contradiction with the downstream dissipation scenarios favored by simultaneous radio and optical observations of blazars during gamma-ray flares \cite{marscher-bllac,fermi-279}. Alternatively, very compact emission regions embedded
within the large-scale jet \cite{ghis-tav,marscher-jorstad}, jet recollimation \cite{nalewajko} or magnetic reconnection events \cite{giannios} could accommodate fast variability with far dissipation scenarios.

\section{VERITAS blazar observations}
VERITAS is an array of four imaging atmospheric Cherenkov telescopes located at the Fred Lawrence Whipple Observatory (FLWO) in southern Arizona ($31^\circ 40'$\,N, $110^\circ 57'$\,W,  1.3\,km a.s.l.). It combines a large effective area over a wide range of energies (100\,GeV to 30\,TeV) with an energy resolution of 15-25\% and an angular resolution of less that $0.1^{\circ}$. The high sensitivity of VERITAS allows the detection of sources with a flux of 0.01 times that of the Crab Nebula in about 25 hours. The standard VERITAS analysis methods are described in \cite{cogan, daniel}.

\begin{table}[t]
\begin{footnotesize}
\begin{center}
\begin{tabular}{c  c  c  c }
\hline
\hline
{\footnotesize Source} & {\footnotesize $z$} &  {\footnotesize Type} & {\footnotesize VERITAS references}\\
\hline
{\footnotesize 3C\,66A$^{\dagger}$} & {\footnotesize 0.444\,?} & {\footnotesize IBL} & ATel~1753, \cite{ver-66a-1, ver-66a-2}\\
{\footnotesize 1ES\,0229+200} & {\footnotesize 0.140} & {\footnotesize HBL} & \cite{icrc-blazars} \\
{\footnotesize RBS\,0413$^{\dagger}$} & {\footnotesize 0.190} & {\footnotesize HBL} & ATel~2272, \cite{icrc-0413}\\
{\footnotesize 1ES\,0414+009} & {\footnotesize 0.287} & {\footnotesize HBL} & \cite{icrc-blazars}\\
{\footnotesize 1ES\,0502+675$^{\dagger}$} & {\footnotesize 0.341\,?} & {\footnotesize HBL} & ATel~2301\\
{\footnotesize VER\,J0521+211$^{\dagger}$} & {\footnotesize ?} & {\footnotesize HBL} & ATel~2260,\,2309, \cite{fermi-ver}\\
{\footnotesize RX\,J0648.7+1516$^{\dagger}$} & {\footnotesize 0.179} & {\footnotesize HBL} & ATel~2486, \cite{ver-0648}\\
{\footnotesize RGB\,J0710+591$^{\dagger}$} & {\footnotesize 0.125} & {\footnotesize HBL} & ATel~1941, \cite{ver-0710}\\
{\footnotesize 1ES\,0806+524$^{\dagger}$} & {\footnotesize 0.138} & {\footnotesize HBL} & ATel~1415, \cite{ver-0806}\\
{\footnotesize Mrk\,421} & {\footnotesize 0.030} & {\footnotesize HBL} & \cite{ver-421,icrc-421} \\
{\footnotesize B2\,1215+30} & {\footnotesize 0.130\,?} & {\footnotesize IBL} & \cite{icrc-blazars}\\
{\footnotesize 1ES\,1218+304} & {\footnotesize 0.184} & {\footnotesize HBL} & \cite{ver-1218-1,ver-1218-2}\\
{\footnotesize W\,Comae$^{\dagger}$} & {\footnotesize 0.102} & {\footnotesize IBL} & ATel~1422,\,1565, \cite{ver-wcom-1, ver-wcom-2}\\
{\footnotesize PKS\,1424+240$^{\dagger}$} & {\footnotesize ?} & {\footnotesize IBL} & ATel~2084, \cite{ver-1424}\\
{\footnotesize H\,1426+428} & {\footnotesize 0.129} & {\footnotesize HBL} & \cite{icrc-blazars}\\
{\footnotesize 1ES\,1440+122$^{\dagger}$} & {\footnotesize 0.162} & {\footnotesize IBL} & ATel~2786, \cite{icrc-discovery}\\
{\footnotesize PG\,1553+113} & {\footnotesize 0.5\,?} & {\footnotesize HBL} & \cite{icrc-1553}\\
{\footnotesize Mrk\,501} & {\footnotesize 0.034} & {\footnotesize HBL} & \cite{ver-501}\\
{\footnotesize 1ES\,1959+650} & {\footnotesize 0.047} & {\footnotesize HBL} & \cite{icrc-blazars}\\
{\footnotesize BL\,Lacertae} & {\footnotesize 0.069} & {\footnotesize LBL} & ATel~3459, \cite{icrc-bllac}\\
{\footnotesize 1ES\,2344+514} & {\footnotesize 0.044} & {\footnotesize HBL} & \cite{ver-2344}\\
\hline\hline
\end{tabular}
\caption{List of the 21 blazars detected at VHE with VERITAS, ordered by right ascension. The 10
VHE discoveries are marked with $\dagger$. Unknown and uncertain redshifts are indicated with a question mark (?). 
The blazar type classifications are taken from \cite{nieppola} or are derived from VERITAS-led multiwavelength studies.}\label{blazar_table}
\end{center}
\vspace{-0.3cm}
\end{footnotesize}
\end{table}

Table~\ref{blazar_table} shows the list of 21 blazars detected in the VHE band with VERITAS since it started regular observations in 2007.  Ten of these detections are first-time discoveries in the VHE band. VERITAS typically accumulates $\sim1100$\,h/year of observations, with blazar observations adding up to $\sim 400$\,h/year \cite{icrc-blazars}. The VERITAS blazar program includes discovery observations, monitoring of known VHE blazars, and target-of-opportunity (ToO) observations of flaring sources. All VERITAS detections are supported by extensive simultaneous multiwavelength observations in the optical, X-ray, and often GeV band, allowing detailed SED modeling to be carried out.

Apart from the positive detections, VERITAS has observed a sample of 47 X-ray/EGRET-selected targets where no significant signal was detected. Although none of these blazars was detected individually, stacking the results from these observations yields an excess with $4.1\sigma$ significance. Interestingly, the same stacking procedure applied to 21 {\em Fermi}-selected blazar targets does not show a positive excess \cite{icrc-discovery}.

VERITAS also has an extensive ToO program to carry out follow-up VHE observations on blazars that are seen in a high state at lower frequencies. VERITAS currently has an existing program using public {\em Fermi}-LAT data to identify high GeV gamma-ray states \cite{icrc-lat}. VERITAS
also responds to public announcements of blazar flares in
the gamma-ray, X ray, and optical bands, and has developed an internal blazar monitoring program at optical frequencies at the UCO/Lick Observatory. A summary of recent results from the ToO program can be found in \cite{icrc-too}.

\begin{figure*}[t]
\centering
\includegraphics[width=0.7\textwidth]{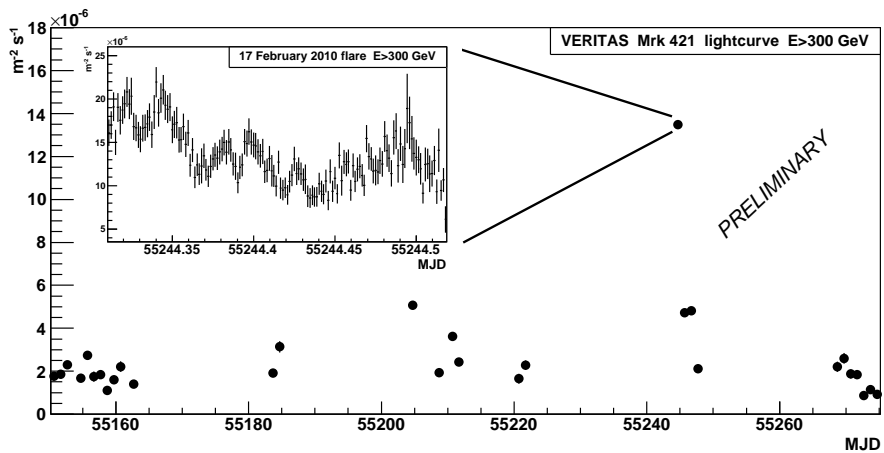}
\caption{Nightly light curve of Markarian~421 during the VERITAS 2009-2010 monitoring campaign.
  The source was detected at an average flux level of approximatively 2 Crab Nebula units over the entire season.
  On  2010 February 17, the source was observed at a flux level of $\sim8$ Crab Nebula units.
  A zoom of the intranight light curve of the flaring event  in 2-minute bins is shown. Taken from \cite{icrc-421}.} 
\label{fig:421}
\end{figure*}

\section{Recent blazar discoveries}
The discovery of new VHE-emitting blazars is one of the science goals of the blazar observing program. Five blazar  discoveries have been reported by the VERITAS collaboration in the last two years.

\noi
{\bf RBS~0413} is an HBL with a redshift of $z = 0.190$. It was a VHE candidate due to its high X-ray flux \cite{costamante}. Deep VERITAS observations were motivated by its hard spectrum in the GeV band measured by {\em Fermi}-LAT \cite{2fgl}.
RBS~0413 was observed by VERITAS 
for $\sim$26\,h of quality-selected live time 
between September 2008 and January 2010.
An excess 
with statistical significance of 5.5$\sigma$ was observed.
The time-averaged spectrum follows a power-law with photon index of $\Gamma = 3.2 \pm 0.7$ and integral flux of  
$\sim$2\% Crab Nebula \cite{icrc-0413}.

\noi
{\bf 1ES~0502+675} observations were motivated by the flux and spectrum 
reported by {\em Fermi}-LAT.
It was observed by VERITAS for $\sim$30 h of 
quality-selected live time between September 2009
and January 2010. A gamma-ray excess corresponding to a statistical significance of $\sim$11$\sigma$ was observed. 
The VHE flux is constant within the observed statistics, with an
average value of $\sim$6\% Crab Nebula, and a power-law spectrum with photon index $\Gamma = 3.92 \pm 0.35$ \cite{icrc-discovery}. 
The reported redshift of 0.341 is uncertain. Following the VERITAS detection, optical spectroscopy was done to determine the redshift value, but no significant absorption or emission lines were seen.

\noi
{\bf VER~J0521+211} was suggested as a VHE candidate following a search carried out inside the VERITAS collaboration for clusters of high-energy photons collected by {\em Fermi}-LAT.
VERITAS observations were taken between 22 and 24 October 2009 (MJD 55126-55128), leading to the detection of a new VHE gamma-ray source: VER~J0521+211 \cite{fermi-ver}. An excess with significance of $5.5\sigma$ was obtained during the first 230 minutes of exposure. VERITAS continued monitoring the new gamma-ray source and detected a high flux state in 22 November 2009 (MJD 55157), when VER~J0521+211 reached a peak flux more than three times as high as during the discovery observations.
A total exposure of 14.5 hours was accumulated, leading to 
a signal with significance of $15.6\sigma$ and power-law spectrum with photon index $\Gamma = 3.44\pm0.20$.
Following the VHE detection, optical spectroscopy observations revealed a continuum-dominated spectrum that identifies the optical counterpart as a BL Lac-type blazar.  No spectral absorption lines could be identified, and therefore the redshift of VER~J0521+211 remains unknown \cite{fermi-ver}.

\noi
{\bf RX~J0648.7+1516} is a ROSAT-detected X-ray source associated with a compact flat-spectrum radio source, located $6.3^\circ$ off  the galactic plane. The source was identified as a promising VHE candidate by the {\em Fermi}-LAT Collaboration, and this information prompted VERITAS observations of the source.
RX~J0648.7+1516 was observed with VERITAS between 4 March and 15 April 2010 (MJD 55259-55301) for a total exposure of 19.3 hours of quality-selected live-time.
A gamma-ray signal with $5.2\sigma$ significance was detected, with a spectrum following a power law with photon index $\Gamma = 4.4\pm0.8$ and integral flux of 7\% that of the Crab Nebula \cite{ver-0648}. 
Optical spectroscopy measurements were triggered following the VERITAS detection. 
The optical spectrum showed Ca H+K, G-band, Na I and Mg I  spectral absorption features, compatible with the source being at a redshift of $z = 0.179$ \cite{ver-0648}.  These observations also provide a definitive optical identification of RX~J0648.7+1516 as a blazar of the BL Lac subclass. 

\noi
{\bf 1ES~1440+122} is an IBL with a redshift of $z = 0.162$. 
This blazar was observed by VERITAS 
for $\sim$47\,h between May 2008 and June 2010.
An excess with statistical significance of 5.5$\sigma$ was detected, corresponding to an integral flux of $\sim$1\% Crab Nebula \cite{icrc-discovery} with photon index $\Gamma = 3.4 \pm 0.7$.

\section{Observations of flaring blazars}


\noi
{\bf Mrk~421} is regularly monitored by VERITAS, 
taking short snapshots at irregular intervals.
On 2010 February 17, VERITAS detected a very bright gamma-ray flare from Mrk~421 and therefore extended the coverage to five hours.
Figure~\ref{fig:421} shows the Mrk~421 nightly
light curve over the entire 2009-10 season, with a detailed view of the 2010 February 17 flaring event.
During the flare, the average gamma-ray flux exceeded 8 Crab Nebula units, with a significance exceeding $260\sigma$ over the 4.9\,h of observation time \cite{icrc-421}. Flux variability is observed on time scales shorter than 10\,min. 

\begin{figure}
\centering
\includegraphics[width=0.99\columnwidth]{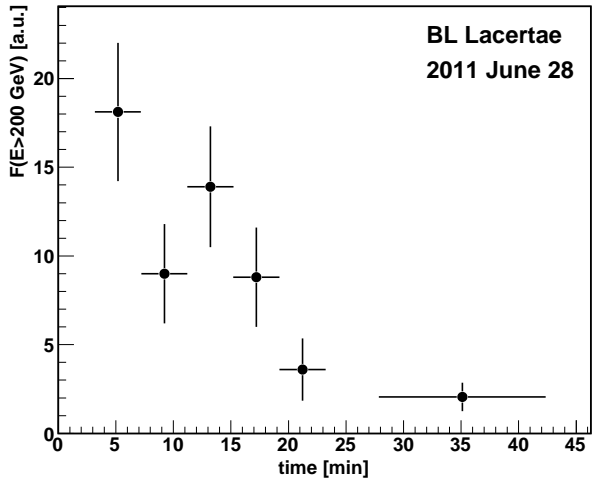}
\caption{Observed gamma-ray light curve of BL~Lacertae on the night of 2011 June 28. Clear flux variability was seen during the 40 minutes of VERITAS observations, with a flux-halving time of $\sim15$\,min. This is the shortest variability timescale observed so far in an LBL. } 
\label{fig:bllac}
\end{figure}

\noi
{\bf BL~Lacertae} was discovered at VHE by MAGIC during a flare in 2005 with a flux of
$\sim$ 3\% Crab Nebula \cite{magic-bllac}. The source has been repeatedly observed with ground-based Cherenkov telescopes since then, but it was not detected again until 2011.
In May 2011, several observatories
reported an increase in activity of the source from optical to GeV energies. VERITAS monitoring observations started  on
2011 May 26 and continued on several nights through the end of June. On 2011 June 28, a flare was observed by VERITAS during a 
40-minute observation under moonlight conditions. 
The observed gamma-ray flux exceeded 50\% Crab Nebula during the flare, with a total significance above $20\sigma$ \cite{icrc-bllac}.
Figure~\ref{fig:bllac} shows the VHE gamma-ray light curve during the observed flare. Clear intranight variability is seen, with halving time of $\sim 15$\,min, which is the fastest variability ever observed in an LBL.

\section{Discussion and conclusions}
The population of VHE-detected blazars is rapidly growing. The VHE blazar sample is dominated by HBLs, but thanks to the increased sensitivity of current ground-based Cherenkov telescopes it also includes a small number of low-frequency-peaked BL~Lacs and FSRQs. VERITAS has contributed to this progress with the discovery of 10 new VHE blazars in the past few years and several key observations of known VHE sources in flaring states, always accompanied by simultaneous multiwavelength coverage.

The VERITAS blazar science program has several goals:
\begin{itemize}
\item Monitoring of a sample of known nearby VHE blazars to study the properties of their gamma-ray emission: flux and spectral variability, duty cycle, extension of the spectrum above 10\,TeV, and measurement of the shortest variability timescales.
\item Extensive observations of known hard-spectrum VHE blazars with the main goal of characterizing their emission above 1\,TeV to study the interaction of multi-TeV photons with the EBL, thereby constraining the EBL density and the magnitude of the primordial intergalactic magnetic fields.
\item Monitoring of bright IBLs and LBLs to characterize their gamma-ray emission during non-flaring episodes and understand under which conditions low-frequency-peaked BL~Lacs emit significant levels of VHE emission.
\item Coordination of broadband multiwavelength observations simultaneous with VERITAS detections of blazars, focusing on radio, optical, X-ray and GeV bands. Multiwavelength coverage is used to build simultaneous SEDs and perform correlated variability studies that can be compared with the predictions from different emission models.
\item Discovery observations of new VHE candidates to enlarge the population of VHE blazars, with special emphasis on increasing the number of detected IBLs, LBLs and FSRQs.
\item A ToO program to follows up on reports of high radio/optical/X-ray/GeV states for selected targets, in order to characterize the VHE emission of these objects and to study correlated flaring behavior between different energy bands.
\end{itemize}

VERITAS will continue to develop a strong program of blazar observations in the coming years.

\bigskip 
\begin{acknowledgments}
This work was supported in part by the NSF grant Phy-0855627 and NASA grant NNX10AP66G at Barnard College. VERITAS research is supported by grants from the US Department of Energy, the US National Science Foundation, and the Smithsonian Institution, by NSERC in Canada, by Science Foundation Ireland, and by STFC in the UK. We acknowledge the excellent work of the technical support staff at the FLWO and at the collaborating institutions in the construction and operation of the instrument.
\end{acknowledgments}

\bigskip 

\end{document}